\documentclass[floatfix,prl,twocolumn,letterpaper,lengthcheck,superscriptaddress,showpacs,amssymb,amsmath,amsfonts,aps,altaffilletter,nofootinbib,nopreprintnumbers,showpacs,longbibliography]{revtex4-1}

\usepackage{color}
\usepackage{hyperref}
\usepackage{amsmath}
\usepackage{multirow}
\usepackage{graphicx}
\usepackage{caption}
\usepackage{subcaption}
\usepackage{placeins}

\newcommand{\tbifurcate}{$t_{\rm bifurcate}\,$}

\newcommand{\ttouch}{$t_{\rm touch}\,$}

\def\insubscript{\rm inner}
\def\outsubscript{\rm outer}
\newcommand{\Sin}{\mathcal{S}_{\insubscript}}

\newcommand{\Sout}{\mathcal{S}_{\outsubscript}}

\begin{document}

\title[]{The interior of a binary black hole merger}

\author{Daniel Pook-Kolb} 
\affiliation{Max-Planck-Institut f\"ur Gravitationsphysik (Albert
  Einstein Institute), Callinstr. 38, 30167 Hannover, Germany}
\affiliation{Leibniz Universit\"at Hannover, 30167 Hannover, Germany}

\author{Ofek Birnholtz}
\affiliation{Center for Computational Relativity and Gravitation,
  Rochester Institute of Technology,
  170 Lomb Memorial Drive, Rochester, New York 14623, USA}

\author{Badri Krishnan} 
\affiliation{Max-Planck-Institut f\"ur Gravitationsphysik (Albert
  Einstein Institute), Callinstr. 38, 30167 Hannover, Germany}
\affiliation{Leibniz Universit\"at Hannover, 30167 Hannover, Germany}

\author{Erik Schnetter}
\affiliation{Perimeter Institute for Theoretical Physics, Waterloo, 
  ON N2L 2Y5, Canada}
\affiliation{Physics \& Astronomy Department, University of Waterloo,
  Waterloo, ON N2L 3G1, Canada}
\affiliation{Center for Computation \& Technology, Louisiana State
  University, Baton Rouge, LA 70803, USA}

\date{2019-03-12}

\begin{abstract}
 
  We find strong numerical evidence for a new phenomenon in a binary
  black hole spacetime, namely the merger of marginally outer trapped
  surfaces (MOTSs).  By simulating the head-on collision of two
  non-spinning unequal mass black holes, we observe that the MOTS
  associated with the final black hole merges with the two initially
  disjoint surfaces corresponding to the two initial black holes. This
  yields a connected sequence of MOTSs interpolating between the
  initial and final state all the way through the non-linear binary
  black hole merger process.  In addition, we show the existence of a
  MOTS with self-intersections formed immediately after the merger.
  This scenario now allows us to track physical quantities (such as
  mass, angular momentum, higher multipoles, and fluxes) across the
  merger, which can be potentially compared with the gravitational
  wave signal in the wave-zone, and with observations by gravitational
  wave detectors. This also suggests a possibility of proving the
  Penrose inequality mathematically for generic astrophysical binary
  back hole configurations.
  
\end{abstract}

\maketitle

The merger of two black holes (BHs) is often visualized by an event
horizon (EH), the boundary of the portion of spacetime causally
disconnected from far away observers.  An example of this is
\cite{Matzner:1995ib}, showing the well known ``pair of pants''
picture of the EH for a binary black hole collision.  However, EHs are
not generally suitable for extracting quantities of physical interest
and tracking them all the way through the merger in quantitative studies.
In perturbative regimes or in cases when the end state of the EH is
known, it is sometimes possible to use EHs to calculate mass, angular
momentum, energy fluxes etc. \cite{Hawking:1972hy}, but this does not
carry over to non-perturbative situations (such as during a binary
black hole (BBH) merger)
\cite{Ashtekar:2004cn,Faraoni:2015pmn,Booth:2005qc,Hayward:2000ca}.

It is much more satisfactory, both for conceptual and practical
reasons, to use instead marginally trapped surfaces, first introduced
by Penrose for proving the singularity theorems \cite{Penrose:1964wq}.
Let $S$ be a closed 2-surface with in- and out-going future-directed
null normals $n^a$ and $\ell^a$ respectively, and let
$\Theta_{(n)}$ and $\Theta_{(\ell)}$ be the corresponding
expansions. Trapped surfaces have $\Theta_{(\ell)}<0$,
$\Theta_{(n)}<0$, while a marginally outer trapped surface (MOTS) has
$\Theta_{(\ell)}=0$ with no restriction on $\Theta_{(n)}$.  The
outermost MOTS on a given Cauchy surface, known as an apparent horizon
(AH), has been used to locate BHs even in the earliest numerical BH
simulations (see e.g. \cite{PhysRevD.14.2443}).  The presence of a
trapped surface in a spacetime shows the presence of a singularity and
an EH. MOTSs were thus used as proxies for EHs which are much harder
to locate numerically.

Over the last two decades, however, it has become clear that MOTSs are
much better behaved than previously expected.  The world tube traced
out by a MOTS during time evolution can be used to study energy
fluxes, the evolution of mass, angular momentum and higher multipole
moments
\cite{Krishnan:2007va,Dreyer:2002mx,Schnetter:2006yt,Gupta:2018znn,Ashtekar:2013qta}. The
world tube can be used as an inner boundary for Hamiltonian
calculations, and the laws of BH mechanics hold
\cite{Ashtekar:2004cn,Booth:2005qc,Gourgoulhon:2005ng,Hayward:2004fz,Jaramillo:2011zw,Krishnan:2013saa,Bousso:2015qqa}. In
general the world tubes can be null, spacelike, timelike, or of mixed
signature
\cite{Ashtekar:2004cn,Bousso:2015qqa,Bousso:2015mqa,Ashtekar:2000sz,Ashtekar:2002ag,Ashtekar:2003hk}.
In stationary spacetimes and in perturbative settings, these
calculations coincide with expectations from EHs, but this framework
is generally applicable.

Despite this progress, there remains a significant gap in our
understanding.  For a BBH merger it is routine to compute physical
quantities for either the two separate initial BHs or for the common
final BH.  It is not clear if there should exist a
relationship between the two regimes separated by the merger.  Neither
is it known whether there is a connected sequence of MOTSs which takes
us from the two separated individual MOTSs to the final one.  The
existence of such a connected sequence would allow the possibility of
tracing physical quantities all the way through the dynamical and
non-linear merger process. These predictions could potentially be
compared with calculations of gravitational wave (GW) signals in the
wave-zone and eventually with observations of GWs, thus offering a
unique probe of dynamical and non-linear gravity.

Another motivation for studying the merger of MOTSs is related to
cosmic censorship and the Penrose inequality.  In 1973, Roger
Penrose proposed an inequality relating the area $A$ of a BH horizon
to the spacetime's total ADM mass $M_{ADM}$ \cite{Penrose:1973um}:
\begin{equation}
  A \leq 16\pi M_{ADM}^2\,.
\end{equation}
As originally formulated by Penrose, this inequality applies to a
marginally trapped surface $S$ formed during gravitational collapse
(though there are examples of AHs which violate the inequality
\cite{BenDov:2004gh}).  A proof of this inequality without using event
horizons is seen as strong support for cosmic censorship.  Thus far,
the inequality has been established rigorously for time symmetric
initial data for an arbitrary number of BHs
\cite{huisken2001,bray2001,Bray:2003ns} (see
\cite{Ludvigsen_1983,Mars:2015iro,Bray:2017phd,Alexakis:2015vma} for
some alternate approaches, and \cite{Mars:2009cj} for a review).  For
a BBH system, the Penrose inequality implies
\begin{equation}
  A_1 + A_2 \leq 16\pi M_{ADM}^2\,,
\end{equation}
where $A_{1,2}$ are the initial areas of the two individual BHs, taken
to be two disjoint MOTSs $\mathcal{S}_1$ and $\mathcal{S}_2$.  Let
$\mathcal{S}_f$ be the final MOTS with area $A_f$.  If there is a
connected sequence of MOTSs which takes us from $\mathcal{S}_{1,2}$ to
$\mathcal{S}_f$, and if $A_1 + A_2 \leq A_f$, then this suggests an
alternative route for a mathematical proof of the inequality for
multiple BHs.

\begin{figure}
  \centering    
  \includegraphics[trim=66 46 53 26,clip,width=\columnwidth]{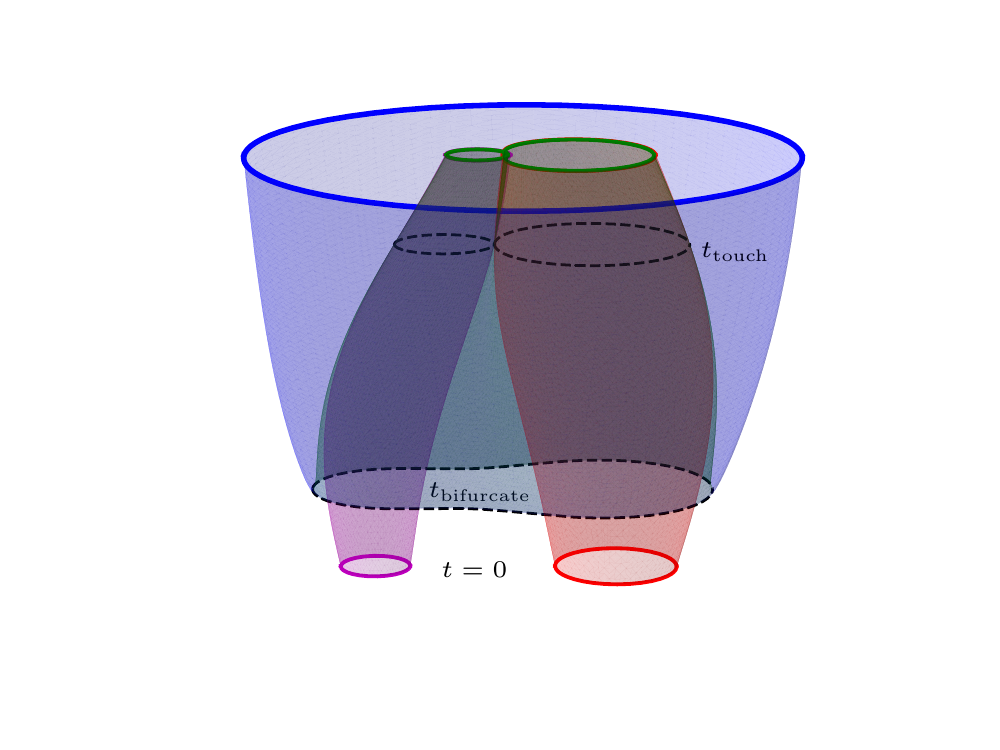}
  \caption{The analog of the \emph{pair-of-pants} picture for MOTSs
    from our numerical simulation.  The tubes traced out by the
    individual MOTSs (colored red and purple) touch and penetrate each
    other. When the individual black holes get sufficiently close a
    common horizon is formed, which bifurcates into an inner branch
    (colored green) and an outer branch (colored blue).  The outer
    branch settles down to the final equilibrium state, while the
    inner branch merges with the individual horizons precisely at the
    time when they touch.}
  \label{fig:3d}
\end{figure}
\emph{Overview of results.}--We address this question by numerically
simulating the head-on collision of two unequal mass black holes. A
rendering of the numerical data from one of our simulations is shown
in Fig.~\ref{fig:3d}; this is the analog of the ``pair of pants''
picture for an event horizon.  The world tubes traced out by the two
individual MOTSs touch at a time \ttouch and then penetrate each
other.  Sometime before \ttouch, at \tbifurcate, the common horizon is
formed and immediately bifurcates into an inner and outer branch.  The
outer branch approaches equilibrium as it loses its asymmetries. In
contrast, the inner branch becomes increasingly distorted and merges
with the individual MOTSs precisely at the moment when they touch.
Interestingly, as shall be detailed below, the inner branch still
continues to exist after this merger, but it develops
self-intersections, thereby providing evidence for topology change.

\begin{figure}    
  \includegraphics[width=\columnwidth]{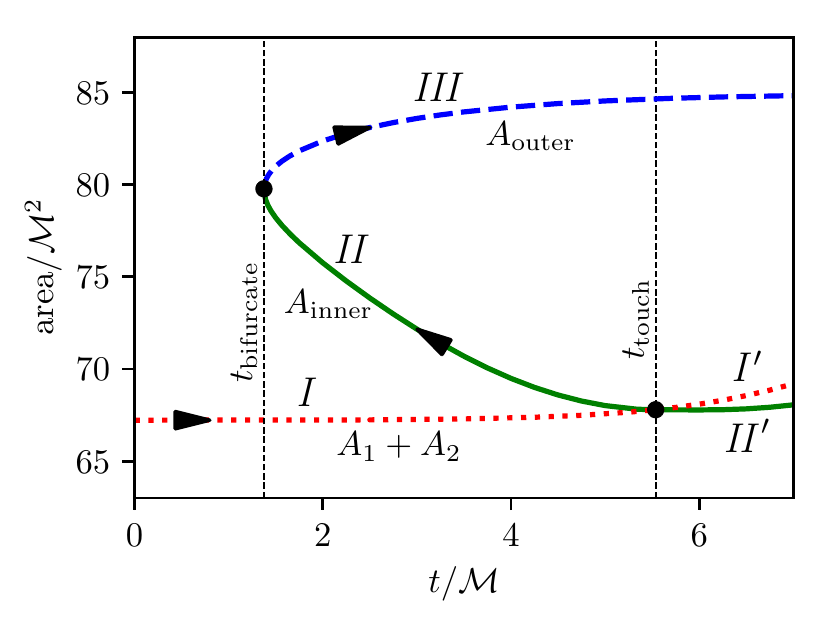}
  \caption{The areas of the various MOTSs as functions of time for the
    same simulation shown in Fig.~\ref{fig:3d}.  The area of the AH is
    shown in blue, the inner common MOTS in green, and the sum of the
    areas of the individual MOTSs in red.  Further details in text.
    Motivated by our choice of parameters, we measure in units of
    $\mathcal{M} = M_{\rm ADM}/1.3$.  }
  \label{fig:bl5-area}
\end{figure}
For the area increase and the Penrose inequality, also to be discussed
further below, we follow the individual MOTSs up to the point when they
touch, and then follow the common MOTS (initially backwards in time),
eventually reaching the final equilibrium state of the outer MOTS.
For the world tubes shown in Fig.~\ref{fig:3d}, this leads to a plot
of the area as a function of time shown in Fig.~\ref{fig:bl5-area}.
We start with the two BHs far apart, represented by the MOTSs
$\mathcal{S}_1$ and $\mathcal{S}_2$, and track their areas $A_1$ and
$A_2$, respectively.  The branch $I$ and $I^\prime$ (dotted red) show
$A_1+A_2$ which is always increasing; $I$ and $I^\prime$ are
respectively the portions before and after \ttouch.

The common horizon is formed with a bifurcation into inner and outer
portions $\Sin$ and $\Sout$ respectively, at the time \tbifurcate\!.
$\Sin$ generates the branch $\mathit{II}$ (solid green), which initially
decreases in area and eventually merges with $I$ at time \ttouch
(which also demarcates $\mathit{II}$ and $\mathit{II}^\prime$).
Segment $\mathit{III}$ (dashed blue) is traced out by the AH which has
increasing area and asymptotes to a final Schwarzschild or Kerr
horizon.  The required sequence of MOTSs is then
$I+\mathit{II}+\mathit{III}$ (segment $\mathit{II}$ is traversed
backwards in time); if we have monotonic area increase along this
sequence, the Penrose inequality is guaranteed to hold.  The portions
$I^\prime$ and $\mathit{II}^\prime$ are not part of this sequence.
Subtleties about the monotonicity of the area will be discussed
further below.

% For a single BH---while we have strong evidence from numerical and
% analytic calculations that the end state of branch $\mathit{III}$ must
% be a Kerr BH---this is not yet mathematically established in the full
% non-linear theory. This work does not add anything in that direction
% but it does suggest a mechanism of extending this result, if it holds,
% to multiple BHs.

\emph{Methodology.}-- Our main technical tool is a new method (and
software) for locating MOTSs numerically which is capable of finding
even very highly distorted MOTSs
\cite{Pook-Kolb:2018igu,pook_kolb_daniel_2019_3352328,Pook-Kolb:2019ssg}.  This is a
modification of the commonly used algorithm known as AHFinderDirect
\cite{Thornburg:1995cp}.  It was previously validated for sequences of
time-symmetric initial data sets, and is here applied during a time
evolution.

We use the Einstein toolkit \cite{Loffler:2011ay, EinsteinToolkit:web}
infrastructure for our calculations. We set up initial conditions via
the two-puncture single-domain method \cite{Ansorg:2004ds} and enforce
axisymmetry following \cite{Pook-Kolb:2019ssg}.  We solve the Einstein
equations in the BSSN formulation as in \cite{Brown:2008sb}, using a
$1+\log$ slicing and a $\Gamma$-driver shift condition, with details
of our initial and gauge conditions as described in
\cite{wardell_barry_2016_155394}.

We use $6$th order finite differencing on a uniform grid spanning
$[0,10] \times [0,0] \times [-10,10]$ and a $6$th order Runge-Kutta
time integrator. Most calculations shown here were performed with a
resolution of $h=1/960$.  Additional resolutions were used to verify
convergence.  All parameter files are available in the repository
\cite{pook_kolb_daniel_2019_3352328}.

%%%%%%%%%%%%%
\begin{figure*}    
  \includegraphics[width=\textwidth]{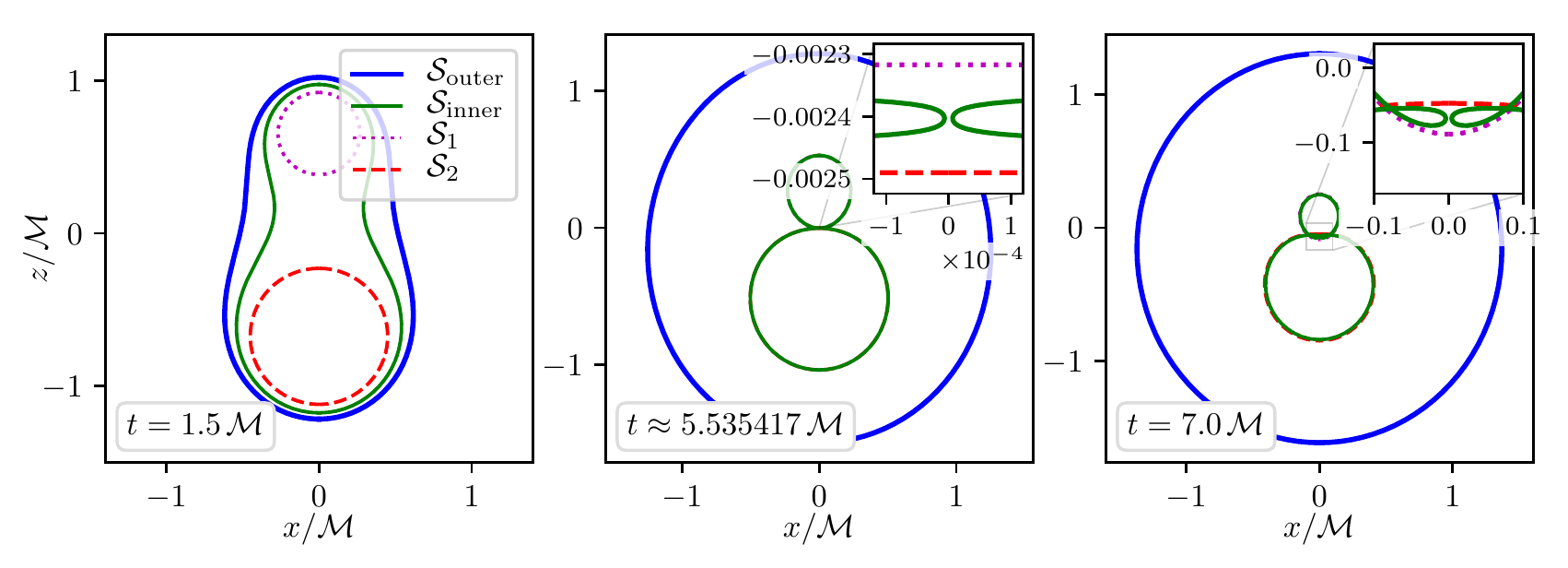}
  \caption{%
    The shapes of the horizons at various times in the simulation;
    this is the same simulation as shown in Figs.~\ref{fig:3d} and
    \ref{fig:bl5-area}.  The numerical values of \ttouch and
    \tbifurcate are found to be \ttouch$\approx 1.374602\,\mathcal{M}$
    and \tbifurcate$\approx 5.537818\,\mathcal{M}$, respectively.  The
    left panel is about $0.1254\,\mathcal{M}$ after \tbifurcate,
    whereas the middle panel is about $0.0024\,\mathcal{M}$ before
    \ttouch.  The right panel is at the end of the simulation, well
    after \ttouch. }
    \label{fig:bl5-overview}
\end{figure*}
%%%%%%%%%%%%%%%%%%%%
\begin{figure}    
    \includegraphics[width=0.45\textwidth]{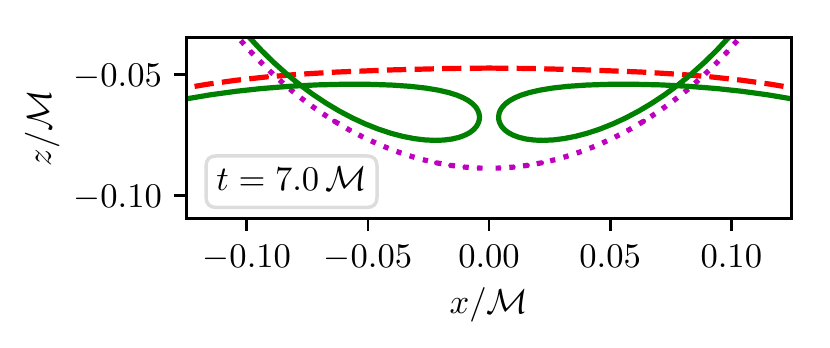}
    \caption{A closer look at the self-intersection for the right
      panel in Fig.~\ref{fig:bl5-overview}. }
    \label{fig:bl5-self-intersect}
\end{figure}
%%%%%%%%%%%%% 
\emph{The merger of the inner horizons.}-- We consider head-on
collisions of non-spinning BHs starting with Brill-Lindquist (BL)
initial data \cite{Brill:1963yv}, representing a BBH system at a
moment of time-symmetry.  The bare masses of the two BHs are denoted
$(m_1,m_2)$ and $d_0$ is the initial separation.  While the Penrose
inequality is known to hold in BL data
\cite{huisken2001,bray2001,Bray:2003ns}, no such time symmetry is
expected to occur in any astrophysical situation in our universe.
Time symmetry implies that the two BHs approach each other and merge
also under time reversal. Furthermore, the incoming radiation at past
null-infinity mirrors the outgoing radiation at future null-infinity.

Some partial results on the behavior of $\Sin$ were known previously
\cite{Gupta:2018znn,Schnetter:2006yt}: $\Sin$ decreases rapidly in
area initially and becomes increasingly distorted as it approaches
$\mathcal{S}_1$ and $\mathcal{S}_2$.  With our new horizon finder, we
are able to track $\Sin$ up to the merger point, and beyond.  We
present our results first for a particular initial configuration
$m_1=0.5$, $m_2= 0.8$ and $d_0 = 1.3$.  We define
$\mathcal{M}=M_{ADM}/(m_1+m_2) = M_{ADM}/1.3$.

We have already shown the world tubes traced out by the MOTSs for this
configuration in Fig.~\ref{fig:3d} and the areas in
Fig.~\ref{fig:bl5-area}.  The shapes of the various marginal surfaces
at selected instants of time are shown in Fig.~\ref{fig:bl5-overview}.
The left panel shows the MOTSs after the AH has formed and $\Sin$ is
fairly distorted.  The center panel shows the MOTSs shortly before
$\mathcal{S}_{1,2}$ touch. The inset shows a close-up of the neck of
$\Sin$, which is very close to pinching off.  The right panel shows
the horizons at a later time when $\mathcal{S}_{1,2}$ penetrate each
other. The penetration of the individual MOTSs was first observed in
\cite{Mosta:2015sga,Szilagyi:2006qy} (see also \cite{Andersson_2009}).
Interestingly, $\Sin$ still continues to exist at this time, but it is
seen to develop self-intersections; Fig.~\ref{fig:bl5-self-intersect}
shows a close-up of the self-intersection.  At later times
$\mathcal{S}_1$ and $\mathcal{S}_2$ continue to move closer and the
``knot'' in $\Sin$ becomes bigger.  We lose numerical resolution at
later times when the horizons get too close to the punctures, and we
have not attempted here to study the eventual fate of the inner
horizons. It is suggested in \cite{Mosta:2015sga} that
$\mathcal{S}_{1,2}$ can cross the punctures and merge, though
discontinuities are observed when the punctures cross the surfaces.
These discontinuities might in fact hide further topology change as
the punctures cross the MOTSs and could perhaps be resolved with our
horizon finder.

% Given that $\Sin$ has a cusp at \ttouch, it is not a
% surprise that we are not able to resolve $\Sin$ very close to the
% merger: this happens in the time interval $\sim 7\times 10^{-3}M$
% before the merger and $\sim 5\times 10^{-3}M$ after it for the
% resolution that we have considered.

%%%%%%%%%%%%%
\begin{figure}
    \includegraphics[width=\columnwidth]{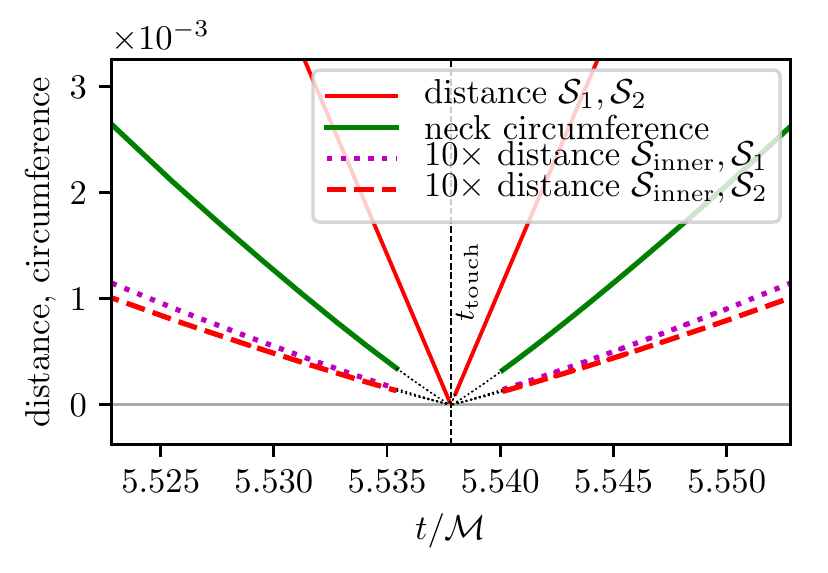}
    \caption{Different measures showing the merger
      $\mathcal{S}_{1,2,\insubscript}$.}
    \label{fig:bl5-merger}
\end{figure}
%%%%%%%%%%%%%
We are led to conjecture that $\Sin$ has a cusp precisely at \ttouch,
and it coincides with $\mathcal{S}_1\cup \mathcal{S}_2$ at this time;
the self-intersections develop immediately after \ttouch.  While
numerical methods will likely not be able to precisely resolve the
instant when the cusp is present, we can provide strong evidence that
the time of the cusp formation coincides with \ttouch.
Fig.~\ref{fig:bl5-merger} shows various quantities which must all
vanish at the point when $\mathcal{S}_{1,2}$ touch.  First, it shows
the proper distance between $\mathcal{S}_1$ and $\mathcal{S}_2$
measured at facing points along the $z$-axis.  Then we plot the proper
circumference of the neck of $\Sin$, and the proper distance between
$\mathcal{S}_{1,2}$ and $\Sin$ along the $z$-axis (the latter
distances are scaled up by a factor of 10 to be properly visible on
this plot).  To define the ``neck'' for a self-intersecting MOTS, we
look at all the curves of rotation obtained by starting with a point
on the knot and rotating it around the symmetry axis.  The neck is the
curve which has smallest proper circumference.  The dotted lines show
the extrapolation to zero whence we see that, as far as we can tell,
$\Sin$ pinches off at the same time (within
$\mathcal{O}(10^{-5})\mathcal{M}$) when $\mathcal{S}_{1,2}$ touch, and
the self-intersections occur immediately after \ttouch.

%%%%%%%%%%%%%
\begin{figure}    
    \includegraphics[width=0.45\textwidth]{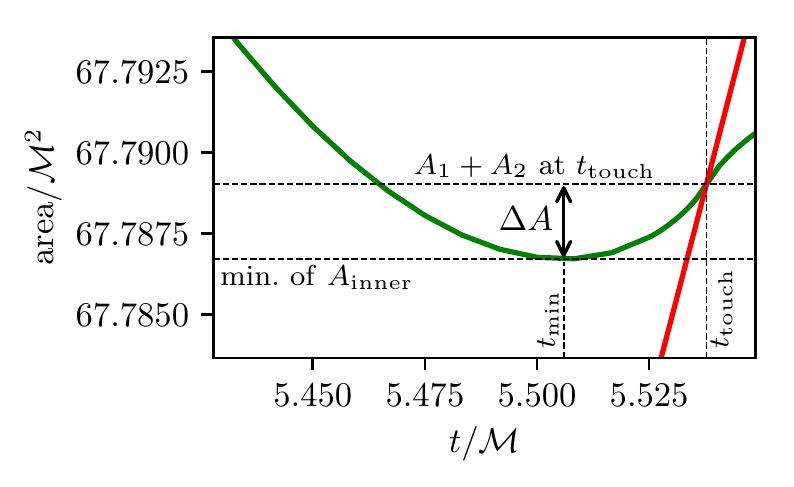}
    \caption{A closer look at Fig.~\ref{fig:bl5-area} near \ttouch
      showing the anomalous area increase of $\Sin$.}
    \label{fig:bl5-area-zoom}
\end{figure}
%%%%%%%%%%%%%
%%%%%%%%%%%%%
\begin{figure}
    \includegraphics[width=0.45\textwidth]{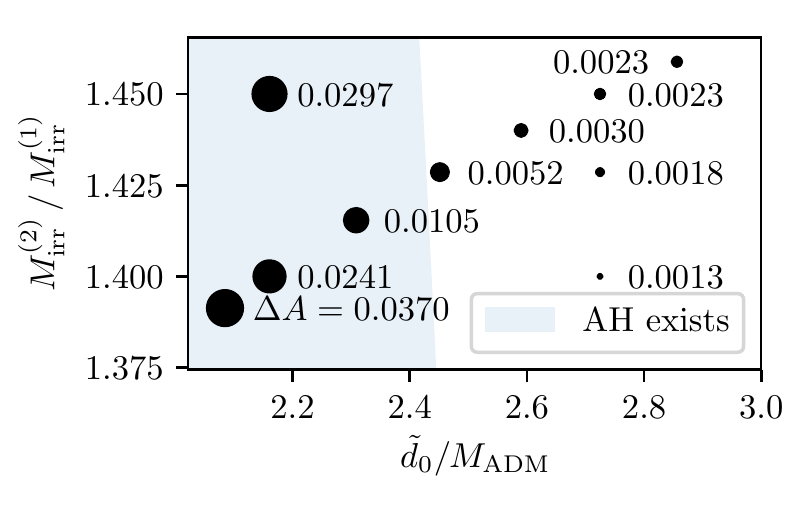}
    \caption{%
      Area increases for various initial conditions. See text for
      details.  }
    \label{fig:area-dips-per-setup}
\end{figure}
%%%%%%%%%%%%%
\emph{The area increase law.}-- From Fig.~\ref{fig:bl5-area}, one
might conclude that the area increases monotonically along the
sequence $\mathit{I}+\mathit{II}+\mathit{III}$ discussed earlier.  A closer look near \ttouch (see
Fig.~\ref{fig:bl5-area-zoom}) reveals a small area \emph{increase}
$\Delta A$ just prior to \ttouch\!; the area is \emph{not} strictly
monotonically increasing along the sequence
$\mathit{I}+\mathit{II}+\mathit{III}$.  The area of course does
increase when we compare the final black hole with the sum of the
initial areas, thus the Penrose inequality is valid.  This result
indicates that a mathematical proof using this route must necessarily
address this behavior, and there might be important physical
information hidden in this area increase.

We repeat the simulations for different mass ratios and initial
separations.  The values of $\Delta A$ over a portion of the parameter
space are shown in Fig.~\ref{fig:area-dips-per-setup}.  The $x$-axis
is the proper distance $\widetilde{d}_0$ between $\mathcal{S}_{1,2}$
in units of the ADM mass, while the $y$-axis is the ratio of the
irreducible masses $M_{irr}^{(1,2)}:= \sqrt{A_{1,2}/16\pi}$ of
$\mathcal{S}_{1,2}$. The values of $\Delta A$ are shown next to the
points and also indicated by the size of the dots.  The configurations
within the shaded region are those which have the common AH already
present in the initial data.  For fixed mass-ratio, $\Delta A$
decreases as $\widetilde{d}_0$ is increased, which suggests that
astrophysical initial data may have vanishing $\Delta A$.

\emph{Conclusions.}-- We have studied the process by which two
marginal surfaces merge to form a common final BH. This is similar to
and complements the ``pair of pants'' picture of a BH merger using
EHs. We have provided strong numerical evidence that there is a
connected sequence of marginal surfaces in this process. This will
potentially allow us to track physical quantities through the merger
and to compare with results obtained from gravitational waveforms.  We
find a new phenomenon, namely the formation of MOTSs with
self-intersections.  Finally, this scenario suggests a different way
of attacking the Penrose inequality. The area increase $\Delta A$ of
$\Sin$ just before the merger is still to be understood and, given the
thermodynamic interpretation of the area, this may contain useful
physical information. 

\emph{Acknowledgments.}-- We thank Lars Andersson, Abhay Ashtekar,
Alex Nielsen, Jeff Winicour, Jose-Luis Jaramillo and the referees for
valuable discussions and suggestions. O.B. acknowledges the National
Science Foundation (NSF) for financial support from Grant
No. PHY-1607520.
This research was also supported by the Perimeter Institute for
Theoretical Physics.  Research at Perimeter Institute is supported in
part by the Government of Canada through the Department of Innovation,
Science and Economic Development Canada and by the Province of Ontario
through the Ministry of Economic Development, Job Creation and Trade.
Some calculations were performed on the \emph{Niagara} cluster of the
University of Toronto.

\bibliography{mots_evolution}{}

\begin{thebibliography}{10}

\bibitem{Alexakis:2015vma}
S.~Alexakis.
\newblock {The Penrose inequality on perturbations of the Schwarzschild
  exterior}.
\newblock 2015.

\bibitem{Andersson_2009}
L.~Andersson, M.~Mars, J.~Metzger, and W.~Simon.
\newblock The time evolution of marginally trapped surfaces.
\newblock {\em Classical and Quantum Gravity}, 26(8):085018, apr 2009.

\bibitem{Ansorg:2004ds}
M.~Ansorg, B.~Br{\"u}gmann, and W.~Tichy.
\newblock A single-domain spectral method for black hole puncture data.
\newblock {\em Phys. Rev. D}, 70:064011, 2004.

\bibitem{Ashtekar:2013qta}
A.~Ashtekar, M.~Campiglia, and S.~Shah.
\newblock {Dynamical Black Holes: Approach to the Final State}.
\newblock {\em Phys. Rev.}, D88(6):064045, 2013.

\bibitem{Ashtekar:2000sz}
A.~Ashtekar et~al.
\newblock {Isolated horizons and their applications}.
\newblock {\em Phys. Rev. Lett.}, 85:3564--3567, 2000.

\bibitem{Ashtekar:2002ag}
A.~Ashtekar and B.~Krishnan.
\newblock {Dynamical horizons: Energy, angular momentum, fluxes and balance
  laws}.
\newblock {\em Phys. Rev. Lett.}, 89:261101, 2002.

\bibitem{Ashtekar:2003hk}
A.~Ashtekar and B.~Krishnan.
\newblock {Dynamical horizons and their properties}.
\newblock {\em Phys. Rev.}, D68:104030, 2003.

\bibitem{Ashtekar:2004cn}
A.~Ashtekar and B.~Krishnan.
\newblock {Isolated and dynamical horizons and their applications}.
\newblock {\em Living Rev. Rel.}, 7:10, 2004.

\bibitem{BenDov:2004gh}
I.~Ben-Dov.
\newblock {The Penrose inequality and apparent horizons}.
\newblock {\em Phys. Rev.}, D70:124031, 2004.

\bibitem{Booth:2005qc}
I.~Booth.
\newblock {Black hole boundaries}.
\newblock {\em Can. J. Phys.}, 83:1073--1099, 2005.

\bibitem{Bousso:2015mqa}
R.~Bousso and N.~Engelhardt.
\newblock {New Area Law in General Relativity}.
\newblock {\em Phys. Rev. Lett.}, 115(8):081301, 2015.

\bibitem{Bousso:2015qqa}
R.~Bousso and N.~Engelhardt.
\newblock {Proof of a New Area Law in General Relativity}.
\newblock {\em Phys. Rev.}, D92(4):044031, 2015.

\bibitem{bray2001}
H.~L. Bray.
\newblock Proof of the riemannian penrose inequality using the positive mass
  theorem.
\newblock {\em J. Differential Geom.}, 59(2):177--267, 10 2001.

\bibitem{Bray:2003ns}
H.~L. Bray and P.~T. Chrusciel.
\newblock {The Penrose inequality}.
\newblock 2003.

\bibitem{Bray:2017phd}
H.~L. Bray and H.~P. Roesch.
\newblock {Null Geometry and the Penrose Conjecture}.
\newblock 2017.

\bibitem{Brill:1963yv}
D.~R. Brill and R.~W. Lindquist.
\newblock {Interaction energy in geometrostatics}.
\newblock {\em Phys. Rev.}, 131:471--476, 1963.

\bibitem{Brown:2008sb}
J.~D. Brown, P.~Diener, O.~Sarbach, E.~Schnetter, and M.~Tiglio.
\newblock {Turduckening black holes: an analytical and computational study}.
\newblock {\em Phys. Rev. D}, 79:044023, 2009.

\bibitem{Dreyer:2002mx}
O.~Dreyer, B.~Krishnan, D.~Shoemaker, and E.~Schnetter.
\newblock {Introduction to Isolated Horizons in Numerical Relativity}.
\newblock {\em Phys. Rev.}, D67:024018, 2003.

\bibitem{EinsteinToolkit:web}
{Einstein Toolkit}: Open software for relativistic astrophysics.
\newblock \url{http://einsteintoolkit.org/}.

\bibitem{Faraoni:2015pmn}
V.~Faraoni and A.~Prain.
\newblock {Understanding dynamical black hole apparent horizons}.
\newblock {\em Lecture Notes in Physics}, 907:1--199, 2015.

\bibitem{Gourgoulhon:2005ng}
E.~Gourgoulhon and J.~L. Jaramillo.
\newblock {A 3+1 perspective on null hypersurfaces and isolated horizons}.
\newblock {\em Phys. Rept.}, 423:159--294, 2006.

\bibitem{Gupta:2018znn}
A.~Gupta, B.~Krishnan, A.~Nielsen, and E.~Schnetter.
\newblock {Dynamics of marginally trapped surfaces in a binary black hole
  merger: Growth and approach to equilibrium}.
\newblock {\em Phys. Rev.}, D97(8):084028, 2018.

\bibitem{Hawking:1972hy}
S.~W. Hawking and J.~B. Hartle.
\newblock {Energy and angular momentum flow into a black hole}.
\newblock {\em Commun. Math. Phys.}, 27:283--290, 1972.

\bibitem{Hayward:2000ca}
S.~A. Hayward.
\newblock {Black holes: New horizons}.
\newblock In {\em {Recent developments in theoretical and experimental general
  relativity, gravitation and relativistic field theories. Proceedings, 9th
  Marcel Grossmann Meeting, MG'9, Rome, Italy, July 2-8, 2000. Pts. A-C}},
  pages 568--580, 2000.

\bibitem{Hayward:2004fz}
S.~A. Hayward.
\newblock {Energy and entropy conservation for dynamical black holes}.
\newblock {\em Phys. Rev.}, D70:104027, 2004.

\bibitem{huisken2001}
G.~Huisken and T.~Ilmanen.
\newblock The inverse mean curvature flow and the riemannian penrose
  inequality.
\newblock {\em J. Differential Geom.}, 59(3):353--437, 11 2001.

\bibitem{Jaramillo:2011zw}
J.~L. Jaramillo.
\newblock {An introduction to local Black Hole horizons in the 3+1 approach to
  General Relativity}.
\newblock {\em Int. J. Mod. Phys.}, D20:2169, 2011.

\bibitem{Krishnan:2007va}
B.~Krishnan.
\newblock {Fundamental properties and applications of quasi-local black hole
  horizons}.
\newblock {\em Class. Quant. Grav.}, 25:114005, 2008.

\bibitem{Krishnan:2013saa}
B.~Krishnan.
\newblock {Quasi-local black hole horizons}.
\newblock In A.~Ashtekar and V.~Petkov, editors, {\em Springer Handbook of
  Spacetime}, pages 527--555. Springer-Verlag, 2014.

\bibitem{Loffler:2011ay}
F.~L{\"{o}}ffler, J.~Faber, E.~Bentivegna, T.~Bode, P.~Diener, R.~Haas,
  I.~Hinder, B.~C. Mundim, C.~D. Ott, E.~Schnetter, G.~Allen, M.~Campanelli,
  and P.~Laguna.
\newblock {{T}he {E}instein {T}oolkit: {A} {C}ommunity {C}omputational
  {I}nfrastructure for {R}elativistic {A}strophysics}.
\newblock {\em Class. Quantum Grav.}, 29(11):115001, 2012.

\bibitem{Ludvigsen_1983}
M.~Ludvigsen and J.~A.~G. Vickers.
\newblock An inequality relating total mass and the area of a trapped surface
  in general relativity.
\newblock {\em Journal of Physics A: Mathematical and General},
  16(14):3349--3353, oct 1983.

\bibitem{Mars:2009cj}
M.~Mars.
\newblock {Present status of the Penrose inequality}.
\newblock {\em Class. Quant. Grav.}, 26:193001, 2009.

\bibitem{Mars:2015iro}
M.~Mars and A.~Soria.
\newblock {On the Penrose inequality along null hypersurfaces}.
\newblock {\em Class. Quant. Grav.}, 33(11):115019, 2016.

\bibitem{Matzner:1995ib}
R.~A. Matzner, H.~E. Seidel, S.~L. Shapiro, L.~Smarr, W.~M. Suen, S.~A.
  Teukolsky, and J.~Winicour.
\newblock {Geometry of a black hole collision}.
\newblock {\em Science}, 270:941--947, 1995.

\bibitem{Mosta:2015sga}
P.~M{\"o}sta, L.~Andersson, J.~Metzger, B.~Szil{\'a}gyi, and J.~Winicour.
\newblock {The Merger of Small and Large Black Holes}.
\newblock {\em Class. Quant. Grav.}, 32(23):235003, 2015.

\bibitem{Penrose:1964wq}
R.~Penrose.
\newblock {Gravitational collapse and space-time singularities}.
\newblock {\em Phys. Rev. Lett.}, 14:57--59, 1965.

\bibitem{Penrose:1973um}
R.~Penrose.
\newblock {Naked singularities}.
\newblock {\em Annals N. Y. Acad. Sci.}, 224:125--134, 1973.

\bibitem{Pook-Kolb:2018igu}
D.~Pook-Kolb, O.~Birnholtz, B.~Krishnan, and E.~Schnetter.
\newblock Existence and stability of marginally trapped surfaces in black-hole
  spacetimes.
\newblock {\em Phys. Rev. D}, 99:064005, Mar 2019.

\bibitem{Pook-Kolb:2019ssg}
D.~Pook-Kolb, O.~Birnholtz, B.~Krishnan, and E.~Schnetter.
\newblock {Self-intersecting marginally outer trapped surfaces}.
\newblock 2019.

\bibitem{pook_kolb_daniel_2019_3352328}
D.~Pook-Kolb, O.~Birnholtz, B.~Krishnan, E.~Schnetter, and V.~Zhang.
\newblock {MOTS Finder version 1.3}, July 2019.

\bibitem{Schnetter:2006yt}
E.~Schnetter, B.~Krishnan, and F.~Beyer.
\newblock {Introduction to dynamical horizons in numerical relativity}.
\newblock {\em Phys. Rev.}, D74:024028, 2006.

\bibitem{PhysRevD.14.2443}
L.~Smarr, A.~\ifmmode \check{C}\else \v{C}\fi{}ade\ifmmode~\check{z}\else
  \v{z}\fi{}, B.~DeWitt, and K.~Eppley.
\newblock Collision of two black holes: Theoretical framework.
\newblock {\em Phys. Rev. D}, 14:2443--2452, Nov 1976.

\bibitem{Szilagyi:2006qy}
B.~Szilagyi, D.~Pollney, L.~Rezzolla, J.~Thornburg, and J.~Winicour.
\newblock {An Explicit harmonic code for black-hole evolution using excision}.
\newblock {\em Class. Quant. Grav.}, 24:S275--S293, 2007.

\bibitem{Thornburg:1995cp}
J.~Thornburg.
\newblock {Finding apparent horizons in numerical relativity}.
\newblock {\em Phys. Rev. D}, 54:4899--4918, 1996.

\bibitem{wardell_barry_2016_155394}
B.~Wardell, I.~Hinder, and E.~Bentivegna.
\newblock {Simulation of GW150914 binary black hole merger using the Einstein
  Toolkit}, Sept. 2016.

\end{thebibliography}
\bibliographystyle{abbrv}

\end{document}